\documentclass[letter,twocolumn]{jpsj3}
\usepackage{txfonts}
\title{Magnetic Excitations of Spin Nematic State in Frustrated Ferromagnetic Chain}

\author{Hiroaki Onishi}
\inst{Advanced Science Research Center, Japan Atomic Energy Agency, Tokai, Ibaraki, 319-1195, Japan}

\abst{%
By exploiting density-matrix renormalization group techniques,
we investigate the dynamical spin structure factor of
a spin-1/2 Heisenberg chain
with ferromagnetic nearest-neighbor
and antiferromagnetic next-nearest-neighbor exchange interactions
in an applied magnetic field.
In a field-induced spin nematic regime,
we find gapless longitudinal and gapped transverse spin excitation spectra,
in accordance with quasi-long-ranged longitudinal
and short-ranged transverse spin correlations, respectively.
The gapless point coincides with the dominant longitudinal spin correlation,
whereas the gap position exhibits a characteristic field dependence
contradicting the dominant transverse spin correlation.
}

\begin{document}
\maketitle


Frustrated quantum magnets are a class of interacting spin systems
in which there are competing interactions that cannot be satisfied simultaneously.
In general,
the combined effects of frustration and quantum fluctuations
prevent conventional magnetic order,
and we envisage the emergence of novel spin states.
Even when the system is magnetically disordered,
we typically find ``hidden'' order described by multiple-spin order parameters,
such as dimer, chiral, and multipole ones.
In this paper, we particularly discuss a quadrupole state,
called a spin nematic (SN) state.
\cite{Andreev1984,Penc-book2011}
A schematic picture of the SN state is that
spins align in a spontaneously chosen axis,
while they still fluctuate within the axis and their directions remain unfixed,
resembling the directional order of rod-shaped molecules in the nematic liquid crystal.

The occurrence of the SN state has been demonstrated
for several kinds of frustrated quantum spin models.
\cite{Chubukov1991,Kecke2007,Vekua2007,Hikihara2008,Sudan2009,Zhitomirsky2010,Sato2013,Shannon2006,Tsunetsugu2006,Lauchli2006,Momoi2012}
Among them,
a spin-$1/2$ chain
with ferromagnetic nearest-neighbor $J_{1}<0$
and antiferromagnetic next-nearest-neighbor $J_{2}>0$ exchange interactions
in a magnetic field $h$,
described by
\begin{equation}
  H =
  J_{1} \sum_{i} \mbox{\boldmath $S$}_{i} \cdot \mbox{\boldmath $S$}_{i+1}
  + J_{2} \sum_{i} \mbox{\boldmath $S$}_{i} \cdot \mbox{\boldmath $S$}_{i+2}
  - h \sum_{i} S_{i}^{z},
\label{eq: H}
\end{equation}
has been attracting much current interest
because of not only the novelty of a theoretically predicted SN state,
\cite{Chubukov1991,Kecke2007,Vekua2007,Hikihara2008,Sudan2009,Zhitomirsky2010,Sato2013}
but also the relevance to a wide variety of edge-shared copper-oxide chain compounds.
\cite{Enderle2010,Masuda2011,Svistov2011,Mourigal2012,Nawa2013,Hase2004,Bush2012,Willenberg2012,Dutton2012,Nawa2014}

On the theoretical side,
the ground-state phase diagram has been studied in detail,
\cite{Hikihara2008,Sudan2009}
and it is now well established that a SN state exists at high fields,
where the quadrupole correlation
$\langle S_{i}^{+}S_{i+1}^{+} S_{j}^{-}S_{j+1}^{-} \rangle$
exhibits a quasi-long-range order
due to the formation of two-magnon bound states.
The longitudinal spin correlation
$\langle S_{i}^{z} S_{j}^{z} \rangle$
also shows a quasi-long-range order,
while the transverse spin correlation
$\langle S_{i}^{+} S_{j}^{-} \rangle$
decays exponentially.
Accordingly, the SN state should have gapless excitations
in the quadrupole and longitudinal spin channels,
whereas there should be a finite excitation gap in the transverse spin channel,
corresponding to the binding energy of a magnon pair.
Note that in the SN regime,
the quadrupole correlation is dominant below the saturation,
and the longitudinal spin-density-wave (SDW) correlation becomes dominant
as the magnetic field is decreased,
although both correlations are quasi-long-ranged.
Thus, hereafter, we refer to this phase as the SN/SDW phase.
When the magnetic field is further decreased,
the ground-state phase changes to a vector chiral (VC) phase at low fields.

Concerning the experimental realization of the SN state
in the fruatrated ferromagnetic chain,
LiCuVO$_{4}$ has been studied frequently
as a prototypical material.
\cite{Enderle2010,Masuda2011,Svistov2011,Mourigal2012,Nawa2013}
In a field-induced phase,
neutron diffraction experiments have found a collinear spin-modulated structure,
which is consistent with theoretical results for
the longitudinal SDW correlation in the SN/SDW phase.
\cite{Masuda2011,Mourigal2012}
The spin correlations exhibit short-range behavior in all directions.
\cite{Mourigal2012}
These observations are indeed suggestive of
the development of the quadrupole correlation.
However,
since the direct observation of the non-magnetic SN state is difficult,
the order parameter is still not identified yet.

To collect evidence that the SN state occurs in reality,
it is important to find any specific indications
in what we observe by using various microscopic probes.
In this context,
recent theoretical studies have pointed out that
the NMR relaxation rate shows characteristic temperature and field dependencies
in the SN/SDW phase.
\cite{Sato2009,Sato2011}
NMR experiments for LiCuVO$_{4}$ have reported consistent results with theoretical predictions,
signaling the formation of the bound magnon pairs.
\cite{Nawa2013}
The properties of the low-energy spin excitation spectra have also been discussed.
\cite{Sato2009,Starykh2014,Onishi2015,Smerald2015}

In this paper,
to clarify the property of the SN/SDW phase
from the viewpoint of the spin dynamics,
we investigate spin excitation spectra in a wide range of momentum and energy
by numerical methods.
We clearly find that the longitudinal spin excitation spectrum is gapless,
while the transverse one is gapped,
as naively expected from the behavior of the spin correlations.
We discuss the field dependence of the spectral weight transfer
in relation with the dominant spin correlation.
A striking feature is that
the momentum position of the gap deviates from
that of the dominant transverse spin correlation
as the system approaches the saturation.


\begin{figure*}[t]
\begin{center}
\includegraphics[width=1.0\linewidth]{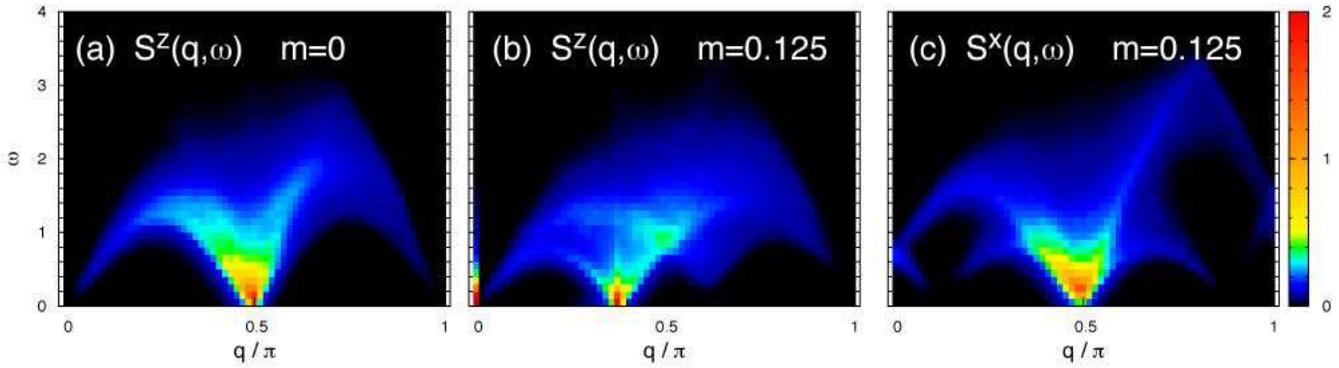}
\end{center}
\caption{(Color online)
Intensity plots of the dynamical spin structure factor $S^{\alpha}(q,\omega)$
at $J_{1}=-1$ and $J_{2}=1$,
obtained by dynamical DMRG calculations with $N=128$.
(a) $S^{z}(q,\omega)$ at $m=0$ and $h=0$.
(b) $S^{z}(q,\omega)$ and (c) $S^{x}(q,\omega)$
at $m=0.125 \, (=16/128)$ and $h=0.649$.
}
\label{fig1}
\end{figure*}

Let us consider the model (\ref{eq: H}) in an $N$-site chain,
and take $J_{2}=1$ as the energy unit.
We investigate the spin excitation dynamics at zero temperature
by exploiting density-matrix renormalization group (DMRG) techniques.
\cite{White1992,Jeckelmann2002}
We employ the finite-system algorithm in open boundary conditions.
We compute the dynamical spin structure factor, defined by
\begin{equation}
  S^{\alpha}(q,\omega) =
  -\frac{1}{\pi}{\rm Im}
  \langle \psi_{\rm G} \vert
  S_{q}^{\alpha \dagger}
  \frac{1}{\omega+E_{\rm G}-H+{\rm i}\eta}
  S_{q}^{\alpha}
  \vert \psi_{\rm G} \rangle,
\end{equation}
where $\vert \psi_{\rm G}\rangle$ is the ground state
with eigenenergy $E_{\rm G}$.
Note that $\vert \psi_{\rm G}\rangle$ is given by the lowest-energy state
in the subspace of a given magnetization $m=M/N$,
where $M=\sum_{i} S_{i}^{z}$.
For the calculation of $S^{\alpha}(q,\omega)$ at $m$,
we set the magnetic field to be
the midpoint of the magnetization plateau of $M$
in the $N$-site system.
$\eta$ is a small broadening factor,
and we set $\eta=0.1$ unless otherwise specified.
Our particular interest is to clarify the anisotropy
between longitudinal $S^{z}(q,\omega)$
and transverse $S^{x}(q,\omega)$
due to the formation of the two-magnon bound states
in the field-induced SN/SDW phase.

For the analysis of the spin excitation spectrum,
we use a dynamical DMRG method,
\cite{Jeckelmann2002}
targeting
the ground state $\vert\psi_{\rm G}\rangle$,
an excitated state $S_{q}^{\alpha} \vert\psi_{\rm G}\rangle$,
and
the so-called correction vector $[\omega+E_{\rm G}-H+{\rm i}\eta]^{-1} S_{q}^{\alpha} \vert\psi_{\rm G}\rangle$.
Here, we note that the truncation error rapidly increases
as the number of target states increases.
Therefore, to obtain $S^{x}(q,\omega)$ with keeping high accuracy,
we calculate $S^{+}(q,\omega)$ and $S^{-}(q,\omega)$ separately,
and then use the relation
$S^{x}(q,\omega)=[S^{+}(q,\omega)+S^{-}(q,\omega)]/4$,
instead of directly calculating $S^{x}(q,\omega)$.
Note also that we calculate the spectrum at $q$ and $\omega$
after one DMRG run with fixed $q$ and $\omega$,
so that we need to perform a great number of DMRG runs
to obtain a full spectrum.


Let us start with a brief discussion on the spectrum at zero field.
In Fig.~1(a),
we present the intensity plot of $S^{z}(q,\omega)$
at $J_{1}=-1$, $J_{2}=1$ (energy unit throughout the paper), $m=0$, and $h=0$.
Note that
$S^{z}(q,\omega)=S^{x}(q,\omega)$ at $h=0$
due to the SU(2) spin rotation symmetry.
We find a sinusoidal dispersion
that gives the lower boundary of a continuum.
The sinusoidal dispersion represents the spinon excitation,
described by the des Cloizeaux-Pearson mode,
\cite{Cloizeaux1962}
since the system decouples into two antiferromagnetic chains
if $J_{1}\rightarrow 0$.
We see a large amount of spectral weight at a lowest-energy peak
$(q_{0},\omega_{0})=(\pi/2,0.04)$.
Note here that the spectrum is asymmetric with respect to $q_{0}$,
as pointed out by the previous studies.
\cite{Enderle2010,Ren2012}
That is, the spectral weight mainly lies near the lower boundary of the continuum
for $q<q_{0}$,
whereas it is distributed to a high energy region for $q>q_{0}$.
On the other hand,
$\omega_{0}$ coincides with a spin excitation energy,
\begin{equation}
  \Delta(N,M) = [E_{0}(N,M+1)+E_{0}(N,M-1)-2E_{0}(N,M)]/2,
\end{equation}
where $E_{0}(N,M)$ is the lowest energy
of the $N$-site system in the subspace of $M$ at $h=0$.
We find that
$\Delta(N,0)$ shifts to lower energy as $N$ increases,
and it is extrapolated to almost zero
in the limit of $N\rightarrow\infty$.
Note that
an exponentially small gap has been predicted by renormalization group theory,
but it is hard to detect such a small gap numerically.
\cite{Itoi2001}

\begin{figure}[b]
\begin{center}
\includegraphics[width=0.95\linewidth]{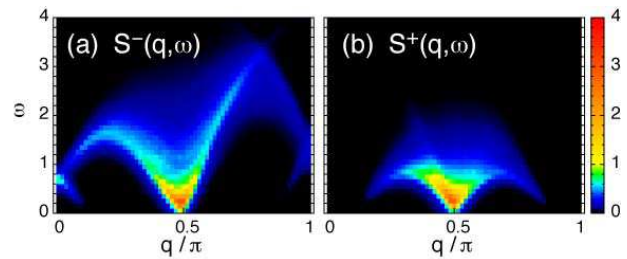}
\end{center}
\caption{(Color online)
Intensity plots of the dynamicsl spin structure factors
(a) $S^{-}(q,\omega)$ and
(b) $S^{+}(q,\omega)$
at $J_{1}=-1$, $J_{2}=1$, $m=0.125$, and $h=0.649$.
Note that $S^{x}(q,\omega)=[S^{+}(q,\omega)+S^{-}(q,\omega)]/4$
is plotted in Fig.~1(c).
}
\label{fig2}
\end{figure}

Now, let us look into the longitudinal and transverse spin excitation spectra
in the SN/SDW phase.
In Figs.~1(b) and 1(c), we show
$S^{z}(q,\omega)$ and $S^{x}(q,\omega)$, respectively,
at $J_{1}=-1$, $m=0.125$, and $h=0.649$.
For $S^{z}(q,\omega)$,
a lowest-energy peak is at
$(q_{0},\omega_{0})=(0.375\pi,0.00)$.
That is, $q_{0}$ moves toward small momentum from the position at zero field.
$\omega_{0}$ is nearly zero,
indicating a gapless mode for the longitudinal spin excitation.
We see that a certain amount of spectral weight is transferred to the origin
due to a finite uniform magnetization,
which is also indicative of a gapless mode.
These gapless points are clearly visible
due to the large intensity.
Moreover,
$S^{z}(q,\omega)$ seems to be gapless at $q=\pi-q_{0}$ and $q=\pi$,
although we do not observe significant intensity
near the possible gapless points in $q>\pi/2$.

In contrast, for $S^{x}(q,\omega)$,
we observe a lowest-energy peak at
$(q_{0},\omega_{0})=(31\pi/64,0.15)$.
That is, $q_{0}$ remains near $\pi/2$ even in the SN/SDW phase.
On the other hand,
$\omega_{0}$ appears to be a finite energy.
In fact,
$\omega_{0}$ agrees with the spin excitation energy $\Delta(N,M)$,
and it is extrapolated to a finite value
in the limit of $N\rightarrow\infty$ with $m=M/N$ fixed,
indicating a gapped mode for the transverse spin excitation.
Note that $S^{x}(q,\omega)$ consists of $S^{-}(q,\omega)$ and $S^{+}(q,\omega)$.
As shown in Fig.~2,
both spectra have
the lowest-energy peak at the same position,
while the overall structure is different between them.
We find that $S^{-}(q,\omega)$ is highly dispersive
in a wide range of momentum and energy,
and $S^{+}(q,\omega)$ is less dispersive
and it is mainly concentrated in a small region
near the lowest-energy peak.

Here, let us discuss how the lowest-energy peak position depends on $m$
in the whole range of $m$.
In Fig.~3(a),
we plot $q_{0}$ of $S^{z}(q,\omega)$
as a function of $m$ for several values of $J_{1}$.
Note that we also find a sharp peak at the origin for finite $m$,
but we focus on the field-induced shift of the peak position
which originally locates near $\pi/2$ at zero field.
At small $m$,
where the system is in the VC phase,
$q_{0}$ shows little dependence on $m$.
At large $m$,
where the system is in the SN/SDW phase,
$q_{0}$ follows the relation $q_{0}=(1/2-m)\pi$
regardless of $J_{1}$,
which supports the bosonization result.
\cite{Sato2009}
Note that $q_{0}$ is represented by the density of bound magnons $1/2-m$.

In Fig.~3(b),
we show the $m$ dependence of $q_{0}$ of $S^{-}(q,\omega)$.
In the VC phase at small $m$,
we notice that $q_{0}$ agrees with the incommensurability
of the transverse spin correlation,
as will be shown in Fig.~4(d).
Note here that
it has been revealed that the incommensurate wave number in the VC phase
is strongly quantum renormalized toward $\pi/2$
compared with the pitch angle $\cos^{-1}(-J_{1}/4J_{2})$
of the helical order in the classical spin case.
\cite{Hikihara2008}
Indeed,
we observe that
$q_{0}=\pi/2$ at $J_{1}=-0.5$ and $-1$,
although the corresponding classical pitch angle is
$0.46\pi$ and $0.42\pi$, respectively.
For small $|J_{1}|$,
we see that $q_{0}$ remains at $\pi/2$
below a threshold value of $m$
even in the SN/SDW phase at large $m$.
However,
$q_{0}$ goes away from $\pi/2$
as $m$ approaches the saturation,
and the deviation from $\pi/2$ is more pronounced
for larger $|J_{1}|$.
On the other hand,
the exact energy dispersion of the one-magnon excited state
in the fully polarized state is
\begin{equation}
  \epsilon_{1}(q)=J_{1}(\cos q-1)+J_{2}(\cos 2q-1)+h,
\label{eq: e1s}
\end{equation}
and its minimum is at
$q=\cos^{-1}(-J_{1}/4J_{2})$,
which coincides with the classical pitch angle.
\cite{Kecke2007}
The present numerical results at the saturation totally agree with
this exact description.
We mention that
the bosonization analysis shows that
the bottom of the one-magnon band is found at $\pi/2$,
\cite{Sato2009}
but it is valid only for the weak-coupling regime $|J_{1}| \ll J_{2}$
and inapplicable in the limit of $m\rightarrow 1/2$.
We thus confirm that
the bottom of the one-magnon band moves
from the quantum renormalized incommensurate wave number
to the classical pitch angle
as $m$ increases from zero to the saturation.
This feature would be useful to determine exchange couplings of real materials.

\begin{figure}[t]
\begin{center}
\includegraphics[width=0.95\linewidth]{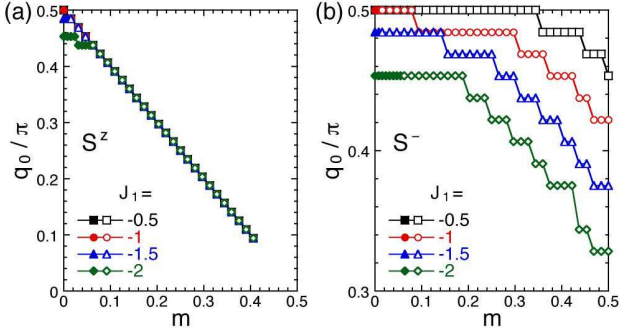}
\end{center}
\caption{(Color online)
The lowest-energy peak position $q_{0}$ of
(a) $S^{z}(q,\omega)$ and
(b) $S^{-}(q,\omega)$
for several values of $J_{1}$ at $J_{2}=1$
as a function of $m$.
Solid symbols denote the VC phase at low fields,
and open symbols denote the SN/SDW phase at high fields.
Note that $q_{0}$ shows a stepwise change simply due to finite-size effects.
The resolution of the momentum is $2\pi/N$ and we use $N=128$ in the present calculations.
Here, the broadning factor is set to $\eta=0.02$
to determine the position of the lowest-energy peak precisely.
}
\label{fig3}
\end{figure}

\begin{figure}[t]
\begin{center}
\includegraphics[width=0.95\linewidth]{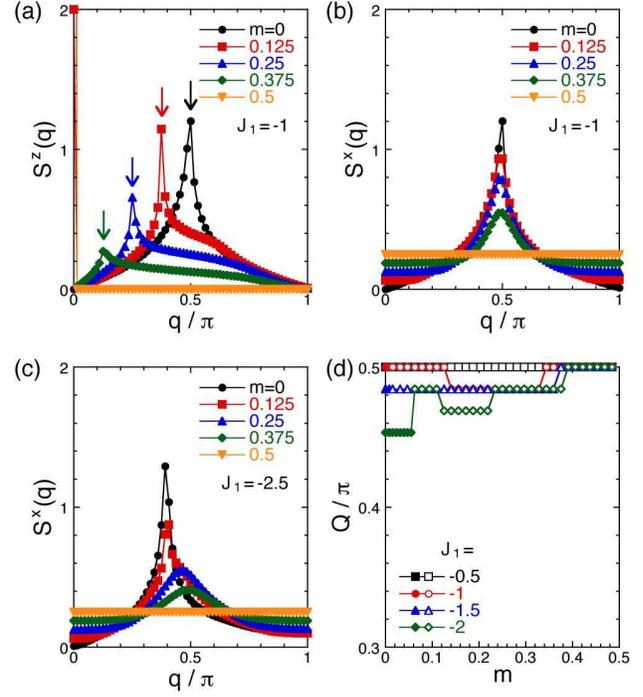}
\end{center}
\caption{(Color online)
(a) $S^{z}(q)$ at $J_{1}=-1$,
(b) $S^{x}(q)$ at $J_{1}=-1$, and
(c) $S^{x}(q)$ at $J_{1}=-2.5$
for several values of $m$.
(d) Pitch angle $Q$,
determined from the maximum point of $S^{-}(q)$,
as a function of $m$.
Solid symbols denote the VC phase at low fields,
and open symbols denote the SN/SDW phase at high fields.
$J_{2}=1$ is fixed.
Data are obtained by DMRG calculations with $N=128$.
}
\label{fig4}
\end{figure}

To gain an insight into the properties of
the field-induced spectral weight transfer
and the dominant spin correlation,
we examine the spin structure factor,
\begin{equation}
  S^{\alpha}(q)=\langle \psi_{\rm G} \vert S_{q}^{\alpha \dagger} S_{q}^{\alpha} \vert \psi_{\rm G} \rangle,
\end{equation}
with attention to sum rules on the integrated intensity of the dynamical spin structure factor.
To obtain $S^{\alpha}(q)$,
we perform ground-state DMRG calculations
independent of dynamical DMRG runs for $S^{\alpha}(q,\omega)$,
since we can obtain accurate results
with relatively small computational cost.
The sum rule for the energy-integrated intensity reads
\begin{equation}
  I^{\alpha}(q) \equiv \int \frac{{\rm d}\omega}{2\pi} S^{\alpha}(q,\omega) = S^{\alpha}(q),
\end{equation}
leading to the spin structure factor.
As for the total intensity
$I^{\alpha} \equiv \sum_{q} S^{\alpha}(q)$,
we have
\begin{equation}
  I^{z}=I^{x}=N/4, \ \
  I^{\pm}=N/2\mp M.
\end{equation}
In Fig.~4(a),
we show $S^{z}(q)$ for several values of $m$ at $J_{1}=-1$.
There is a clear peak at $q=\pi/2$ for $m=0$,
and it shifts toward small momentum with increasing $m$,
as indicated by vertical arrows.
The peak position of $S^{z}(q)$ representing the SDW correlation
is in agreement with the lowest-energy peak position $q_{0}$ of $S^{z}(q,\omega)$
in Fig.~3(a).
We also find that $S^{z}(q)$ grows at $q=0$ as $m$ increases.
In other momentum parts,
$S^{z}(q)$ is suppressed so as to keep the total intensity $I^{z}=N/4$.
Accordingly,
$S^{z}(q,\omega)$ exhibits the spectral weight transfer,
as seen in Fig.~1(b).

We show $S^{x}(q)$ at $J_{1}=-1$ and $J_{1}=-2.5$
in Figs.~4(b) and 4(c), respectively.
We find a sharp peak structure
in the VC phase at small $m$,
while
the peak structure becomes broad
in the SN/SDW phase at large $m$.
This is because
the transverse spin correlation exhibits an algebraic decay
in the VC phase,
and it is short-ranged
in the SN/SDW phase.
As $m$ increases,
the borad peak is leveled off, while the baseline goes up,
and eventually we find a completely flat profile,
i.e., $S^{x}(q)=1/4$, at the saturation.
On the other hand,
the peak position depends on $J_{1}$ and $m$.
Indeed, we find a peak near $q=\pi/2$ regardless of $m$
for $J_{1}=-1$
[Fig.~4(b)],
while
a peak is at $q=25\pi/64$ at $m=0$
and it moves to $q=\pi/2$ near the saturation
for $J_{1}=-2.5$
[Fig.~4(c)].

To clarify the $m$ dependence of the dominant $q$ component
of the transverse spin correlation,
we determine a pitch angle $Q$
from the maximum point of $S^{-}(q)$,
as shown in Fig.~4(d).
Note that
$S^{-}(q)$ and $S^{+}(q)$ have a peak at the same position.
In the VC phase at small $m$,
$Q$ agrees with
the lowest-energy peak position $q_{0}$ of $S^{-}(q,\omega)$, 
as denoted by solid symbols in Figs.~3(b) and 4(d).
However,
in the SN/SDW phase at large $m$,
$Q$ behaves quite differently from $q_{0}$ as $m$ approaches the saturation.
We find that $Q$ changes toward $\pi/2$ regardless of $J_{1}$,
rather than the classical pitch angle $\cos^{-1}(-J_{1}/4J_{2})$ as $q_{0}$,
indicating that the dominant $q$ component is not equivalent to
the momentum position of the opening gap.
This discrepancy is naturally understood
in terms of the short-range nature of the transverse spin correlation.
It is inadequate to take the only leading asymptotic term
that represents the exponential decay
of the transverse spin correlation function
in order to describe the excitation dynamics correctly.
In fact, 
we have seen that
$S^{x}(q)$ turns into the flat profile at the saturation,
meaning that not only the dominant $q$ component but also other components
equally contribute to the spectral weight.

\begin{figure}[t]
\begin{center}
\includegraphics[width=0.6\linewidth]{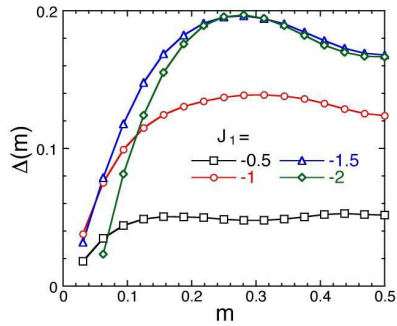}
\end{center}
\caption{(Color online)
The extrapolated gap $\Delta(m)$
for several values of $J_{1}$ at $J_{2}=1$
as a function of $m$.
Here we plot data in the SN/SDW phase.
Note that the zero-field gap is tiny,
\cite{Itoi2001}
and the VC phase is supposed to be gapless.
}
\label{fig5}
\end{figure}

Finally,
we present the dependence of the energy gap of the transverse spin excitation in Fig.~5.
We extrapolate finite-size data to the thermodynamic limit $N\rightarrow\infty$
by assuming a linear relation $\Delta(N,M)=\Delta(m)+a/N$.
For $m \gtrsim 0.1$,
we find that $\Delta(m)$ increases as $|J_{1}|$ becomes large for $J_{1}=-0.5$, $-1$, and $-1.5$,
indicating that the ferromagnetic exchange interaction stabilizes the two-magnon bound state.
$\Delta(m)$ turns to decrease with further increasing $|J_{1}|$,
shown for $J_{1}=-2$.
For small $m$,
$\Delta(m)$ is rapidly reduced as $m$ decreases down to the boundary
between the SN/SDW and VC phases.


In summary,
we have studied the spin excitation dynamics
of the frustrated ferromagnetic chain in the magnetic field
by numerical methods.
In the field-induced SN regime,
the spin excitation spectra exhibit highly anisotropic behavior
between gapless longitudinal and gapped transverse compoments.
The field dependence of the gapless point of $S^{z}(q,\omega)$ is
consistent with the dominant longitudinal spin correlation.
In contrast,
the gap position of $S^{x}(q,\omega)$ shows a unique field dependence
that contradicts with the dominant transverse spin correlation.
We hope that these features could be examined
by inelastic neutron scattering experiments
when searching for signatures of the SN/SDW phase.
On the other hand,
we should observe gapless excitations in the quadrupole channel.
It would be an interesting future problem to study
the quadrupole excitation dynamics and its possible relevance to observables.
\cite{Voll2015}


\begin{acknowledgment}
The author thanks T. Masuda, S. Maekawa, M. Mori, and T. Sugimoto
for discussions and comments.
Part of computations were carried out on the supercomputers
at the Japan Atomic Energy Agency
and the Institute for Solid State Physics, the University of Tokyo.
\end{acknowledgment}


\end{document}